# Combinatorial synthesis and characterization of thin film $Al_{1-x}RE_xN$ ($RE$ = $Pr^{3+}$ and $Tb^{3+}$) heterostructural alloys

Binod Paudel, *[a,b] John S. Mangum,[b] Christopher L. Rom,[b] Kingsley Egbo,[b] Cheng-Wei Lee,[a,b] Harvey Guthrey,[b] Sean Allen,[b] Nancy M. Haegel,[b] Keisuke Yazawa,[a,b] Geoff L. Brennecka[b] and Rebecca W. Smaha *[b]

[a] *Department of Metallurgical and Materials Engineering, Colorado School of Mines, Golden, CO 80401, USA. E-mail: binod.paudel@mines.edu*

[b] *Materials Science Center, National Renewable Energy Laboratory, Golden, CO 80401, USA Address here. E-mail: Rebecca.Smaha@NREL.gov*

**Abstract**

The potential impact of cation-substituted AlN-based materials, such as $Al_{1-x}Sc_xN$, $Al_{1-x}Ga_xN$, and $Al_{1-x}B_xN$, with exceptional electronic, electromechanical, and dielectric properties has spurred research into this broad family of materials. Rare earth (RE) cations are particularly appealing as they could additionally impart optoelectronic or magnetic functionality. However, success in incorporating a significant level of RE cations into AlN has been limited so far because it is thermodynamically challenging to stabilize such heterostructural alloys. Using combinatorial co-sputtering, we synthesized $Al_{1-x}RE_xN$ ($RE$ = Pr, Tb) thin films and performed a rapid survey of the composition-structure-property relationships as a function of RE alloying. Under our growth conditions, we observe that $Al_{1-x}Pr_xN$ maintains a phase-pure wurtzite structure until transitioning to amorphous for $x \gtrsim 0.22$. $Al_{1-x}Tb_xN$ exhibits a phase-pure wurtzite structure until $x \lesssim 0.15$, then exhibits mixed wurtzite and rocksalt phases for $0.16 \lesssim x \lesssim 0.28$, and finally becomes amorphous beyond that. Ellipsometry measurements reveal that the absorption onset decreases with increasing rare earth incorporation and has a strong dependence on the phases present. We observe the characteristic cathodoluminescence emission of $Pr^{3+}$ and $Tb^{3+}$, respectively. Using this synthesis approach, we have demonstrated incorporation of Pr and Tb into the AlN wurtzite structure up to higher compositions levels than previously reported and made the first measurements of corresponding structural and optoelectronic properties.

**Introduction**

Wurtzite-structured (wz) III-nitrides such as aluminum nitride (AlN) and gallium nitride (GaN) have garnered significant attention as next-generation semiconductors thanks to their wide bandgaps (6.4 eV and 3.4 eV, respectively), high mobilities, and exceptional thermal properties. These materials show great promise for high voltage - high frequency power electronics and superior opto(electronic) applications[1,2], including high mobility electron transistors (HEMT)[3] and LEDs. AlN is also indispensable in micro-electromechanical systems (MEMS)[4] and radio frequency communications for acoustic resonators and filter devices[5], and significant performance enhancements have been realized via alloying with ScN.[6,7] This has driven increased research interest in creating and understanding heterostructural alloys (i.e., $III_{1-x}M_xN$, $M$ = metals) between wurtzite AlN and end members that are stable in competing non-wurtzite $M$N phases[8]. This interest accelerated in the last five years with the first report of wurtzite ferroelectricity in $Al_{1-x}Sc_xN$[9], and later $Zn_{1-x}Mg_xO$[10], $Al_{1-x}B_xN$[11], $Ga_{1-x}Sc_xN$[12], and $Al_{1-x}Y_xN$[13].

The rare earth (RE) elements are particularly interesting as alloying options because of the potential for introducing specific magnetic and/or optoelectronic characteristics not typically found in III-nitrides.[14,15] Recent calculations on Gd-, Tb-, and Pr-substituted AlN and GaN alloys predicted that the wurtzite structure remains the lowest-energy crystalline phase across a broad compositional space,[16] but the inherent thermodynamic instability of the alloy requires kinetic control to form and maintain single-phase samples. Synthesis and characterization of $Al_{1-x}RE_xN$ alloys across the $RE^{3+}$ cation series thus remains very limited.

Previous studies attempting to incorporate trivalent REs into a wurtzite AlN lattice had limited and often narrow compositional ranges[14,17–19]. Many such studies used ion implantation of the REs into AlN films initially fabricated by sputtering, MBE, or MOCVD [20–24] followed by post-implantation annealing. The highest concentrations of Tb in AlN with pure phase previously reported were <1 at.% in a powder AlN:Tb phosphor that also showed evidence of an unidentified secondary phase[19,25] and 11 at.% (i.e., $x$≈0.11) in thin films deposited from DC sputtering, though the crystalline phase of these films is unknown[26]. As such, a fundamental understanding of the composition-phase-property relationship(s) within this family is lacking. High throughput combinatorial synthesis is a powerful tool to quickly survey broader phase spaces and perform initial systematic studies of structure(s) and properties across the composition range. Our previous work successfully explored one member of this family by demonstrating the incorporation of $Gd^{3+}$ cations ions up to ~24% in wz-AlN. DFT calculations predicted similar mixing enthalpies for wz-$Al_{1-x}RE_xN$ ($RE$ = Gd, Pr, Tb), suggesting that experimental incorporation is likely achievable. Exploring similarly high substitution levels with other $RE^{3+}$ cations will further expand the current understanding of the effects of ionicity, size mismatches, and chemistries in $Al_{1-x}RE_xN$ films. Optoelectronic properties are of particular interest for $Al_{1-x}RE_xN$ thin films, as the large bandgap of the AlN host lattice offers a wide range of application options for utilizing the sharp well-defined emissions from RE internal 4$f$ transitions[27] spanning UV to IR regions[28–34]. In addition, given the intriguing magnetic ground states in rocksalt (rs) structured-$RE$Ns such as ferromagnetism in GdN and TbN[35,36], studying their alloys with wz-AlN could potentially unlock interesting magnetic phenomena.

Here, we report the combinatorial synthesis of $Al_{1-x}RE_xN$ ($RE$ = Pr, Tb) thin films via radio frequency (RF) co-sputtering. Common synthesis conditions were employed to enable comparison between the RE cations. This yielded films with a broad range of $x$, allowing us to perform a rapid, systematic exploration of the phase space and investigate property trends across the composition range. Scanning/transmission electron microscope (S/TEM) imaging with energy-dispersive X-ray spectroscopy (EDS) mapping reveals polycrystalline films with uniform RE incorporation. We report the growth of single-phase wurtzite $Al_{1-x}RE_xN$ ($RE$ = Pr, Tb) films up to $x \lesssim 0.22$ and $x \lesssim 0.15$, respectively, despite the large ionic size mismatches. $Al_{1-x}Tb_xN$ phase separates into wz+rs phases for $0.16 \lesssim x \lesssim 0.28$ and is amorphous for higher RE content ($x \gtrsim 0.28$), while $Al_{1-x}Pr_xN$ films have a direct crossover from crystalline wz to amorphous at $x \gtrsim 0.22$ with no evidence of any crystalline rs phase. Spectroscopic ellipsometry shows an expected decrease of the absorption onset on the amount of RE incorporation and the phases present. Cathodoluminescence (CL) measurements further confirm the $RE^{3+}$ incorporation within the films and suggest that incorporation of the RE emitters directly into the wz host lattice may provide additional ways to control the ratio of radiative to non-radiative recombination among the various emission lines or via co-doping strategies. This work reports wz-$Al_{1-x}RE_xN$ ($RE$ = Pr, Tb) films up to the highest levels of RE incorporation reported to date with associated data on absorption and emission spectra across composition and phase space.

**Experimental**

**Synthesis**

Combinatorial thin films of Al$_{1-x}$RE$_x$N (RE= Pr and Tb) were synthesized using radio frequency co-sputtering of metallic Al (99.999%, Kurt J. Lesker) and Pr or Tb (99.9%, QS Advanced Materials) targets from 2" magnetrons within a sputtering chamber with a base pressure of ~8×10$^{-8}$ Torr. The magnetrons were in a standard balanced magnetron configuration and placed at a 180° angle to each other. Films were grown on 5.08 × 5.08 cm p-type Si(100) (pSi) substrates with native SiO$_2$ layers that were heated to approximately 600 °C. A cryogenic sheath was employed to trap adventitious oxygen or water during deposition. The targets were pre-sputtered for 45–120 minutes with the substrate shutter closed before depositing for 120 minutes. The Al target power was kept constant at 150 W for all depositions, while powers to RE targets were 20, 40, and 60 W. All films were deposited at a constant gas ratio of Ar:N$_2$ = 2:1 (40:20 sccm) at a process pressure of 4 mTorr. After the depositions, the substrates were cooled in a N$_2$ environment at a pressure of 8 mTorr until the substrate temperature reached <100 °C.

**Characterization**

Experimental data of the combinatorial films were analyzed using the COMBIgor software package[37] and are publicly available in the National Renewable Energy Laboratory (NREL) high throughput experimental materials database[38,39].

Cation composition was measured with X-ray fluorescence (XRF) using a Bruker M4 Tornado under vacuum (~15 Torr). Thicknesses were extracted from the XRF data following calibration of the model with transmission electron microscopy (TEM). The average elemental composition was analyzed at the University of Oregon Center for Advanced Materials Characterization in Oregon (CAMCOR) using a Cameca SX100 electron probe microanalyzer (EPMA) with wavelength dispersive spectrometry (WDS). All sites on samples and standards were measured with a 5 µm wide beam with 30 nA of current at incident electron beam energies of 10, 15, and 20 keV. Raw k-ratios were then imported into STRATAGem thin film processing software to quantify thin film compositions.

S/TEM high-angle annular dark-field (HAADF) and selected area electron diffraction (SAED) images were acquired with a Thermo Fisher Scientific Spectra 200 transmission electron microscope operating at an accelerating voltage of 200 kV. Specimens for TEM were prepared from deposited films via in situ focused ion beam lift-out methods[40] using an FEI Helios Nanolab 600i SEM/FIB DualBeam workstation. Chemical mapping was performed in the TEM using the Super-X energy-dispersive X-ray spectroscopy (EDS) system equipped with four windowless silicon drift detectors, allowing for high count rates and chemical sensitivity (down to 0.5–1 at%). The EDS data were quantified using a multi-polynomial parabolic background and absorption correction in Velox.

Initial structure characterization of the films was performed with laboratory X-ray diffraction using a Bruker D8 Discover with Cu K$_\alpha$ radiation equipped with a two-dimensional (2D) detector. Synchrotron grazing incidence wide angle X-ray scattering (GIWAXS) measurements were performed on selected samples at beamline 11-3 at the Stanford Synchrotron Radiation Lightsource, SLAC National Accelerator Laboratory. The data were collected with a Rayonix 225 area detector at room temperature using a wavelength of λ=0.97625Å, a 1° incident angle, a 150 mm sample-to-detector distance, and a beam size of approximately 50 μm vertical × 150 μm horizonal. The detector images were calibrated with a LaB$_6$ standard, integrated with the Nika SAS package using Igor Pro[41], and processed with PyFAI and pygix.[42,43] Integrated data were averaged from 5 frames of 15 seconds each. LeBail fits of integrated plots were performed using Topas software by refining lattice parameters, sample displacements, and crystallite sizes[44], and obvious outlier peaks from hot pixels were smoothed out manually.

Spectroscopic ellipsometry data were acquired at 65°, 70°, and 75° incident angles on a single row of all sample libraries using a J.A. Woollam Co. M-2000 variable angle ellipsometer. Complete EASE software was used to model the data by fitting the real and imaginary parts of the dielectric function with a four-layer model consisting of the silicon substrate, native silicon oxide, the Al$_{1-x}$RE$_x$N films, and surface roughness approximated with a standard mixed film/void Bruggeman EMA layer. The silicon and native oxide were modeled using well-known optical constants provided by the Complete EASE software. All the Al$_{1-x}$RE$_x$N film layers were modeled using Tauc-Lorentz and Cody-Lorentz semiconductor oscillator models which are established models for accurately fitting the imaginary part of the dielectric function of polycrystalline and/or amorphous materials while maintaining the Kramers−Kronig consistency.

Cathodoluminescence (CL) spectra were collected at room temperature on a JEOL JSM-7600 FESEM equipped with a Horiba H-CLUE CL system. An accelerating voltage of 5kV and incident beam current of ~ 3 nA were used to excite excess carriers in the samples. CL spectra were acquired by scanning from 200 – 1150 nm with an integration time of 3 s, and three acquisitions were averaged to improve the signal-to-noise ratio.

**Calculations**

We calculated the variation in bandgap as a function of composition in the $Al_{1-x}RE_xN$ ($RE$ = Pr and Tb; $x$ = 0.042, 0.125, and 0.250) alloys using the special quasi-random structure (SQS) approach[45] and chose 48-atom supercells to have manageable computational cost for GW calculations. We constructed the supercells using the Alloy Theoretic Automated Toolkit (ATAT)[46] and considered all the pair and triplet clusters within the maximal range of 4.5 and 3.2 Å, respectively. We used Vienna Ab-initio Simulation Package (VASP 5.4.4) to calculate electronic structure. We fully relaxed the SQS supercells (ionic positions, cell size, and shapes) with the cutoff total energy difference of $10^{-5}$ eV. For all the DFT-based calculations, we used Perdew-Burke-Ernzerhof (PBE) exchange-correlation functional. The planewave cutoff kinetic energy was set at 340 eV and self-consistency was ensured with a cutoff total energy of $10^{-6}$ eV. The Brillouin zone was sampled using automatically generated $\mathbf{\Gamma}$-centered Monkhorst-Pack **k**-point grids with a length of 20 for structural relaxation. For GW calculations, we used $\mathbf{\Gamma}$-centered 2x2x2 **k**-point grids and used the PBE wavefunction of the optimized structures as the initial wavefunction. We fixed the wavefunction and iterated the eigenvalues to achieve self-consistency ($\Delta E_g < 0.1$ eV). The number of unoccupied bands was set to be at least ten times the number of occupied bands to obtain converged GW eigen energies.

**Results and discussion,**

**Synthesis and Composition**

Using reactive RF co-sputtering, we synthesized combinatorial $Al_{1-x}RE_xN$ ($RE$ = Pr, Tb) films on pSi (100) substrates intentionally making a one-dimensional (1D) compositional gradient. The co-sputtering schematic is shown in Figure S1. Before depositing the samples with composition gradients, initial growth optimization was performed for crystalline $Al_{1-x}Tb_xN$ films by varying substrate temperatures (400–700 °C), gas ratios (N$_2$:Ar = 3:1–1:3), and Tb target powers (30–75 W). The deposition parameters chosen (see Methods) were then used to grow the final series of Al$_{1-x}$Tb$_x$N and Al$_{1-x}$Pr$_x$N films. We note that the deposition parameters used here are not the same as in our previous work on Al$_{1-x}$Gd$_x$N,[16] which employed a more nitrogen-rich process environment.

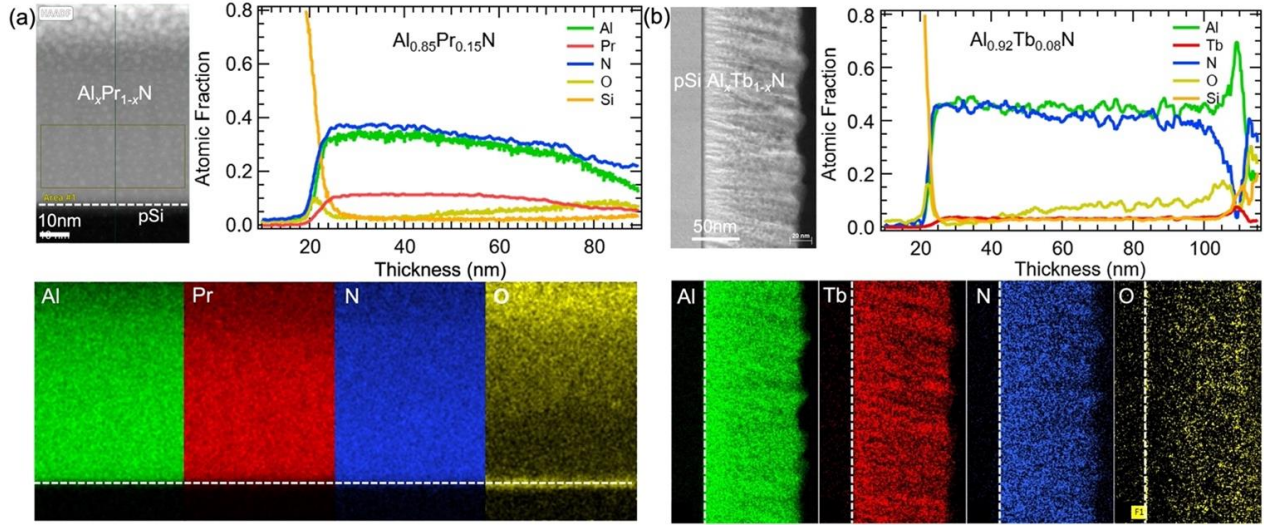

*Figure 1: STEM-HAADF image; EDS elemental mapping at the Al K-peak, Tb or Pr L-peak, N K-peak, and O K peak; and EDS atomic % line profiles of (a) Al$_{0.85}$Pr$_{0.15}$N and (b) Al$_{0.92}$Tb$_{0.08}$N films on pSi substrates.*

The composition of the films was measured with electron probe microanalysis (EPMA) across the compositional gradients, yielding RE compositions of 0.04≲$x$≲0.32 for Al$_{1-x}$Pr$_x$N and 0.08≲$x$≲0.45 for Al$_{1-x}$Tb$_x$N. The nitrogen content ratio N/(N+O), also probed by EPMA, is presented in Figure S2; the highest N/(N+O) ratio was 0.90 for Al$_{1-x}$Pr$_x$N films, although it decreased to ~0.68 near $x$≈0.1. The average value for Al$_{1-x}$Tb$_x$N is higher, approximately 0.87.

To investigate the microstructure, composition, and uniformity of these materials, we investigated with STEM select films whose compositions as measured by EPMA were Al$_{0.84}$Pr$_{0.14}$N$_{0.96}$O$_{0.25}$ (denoted "Al$_{0.85}$Pr$_{0.15}$N") and Al$_{0.91}$Tb$_{0.07}$N$_{0.93}$O$_{0.04}$ (denoted "Al$_{0.92}$Tb$_{0.08}$N"); see Table S1. STEM-HAADF, (Figure 1 top left panels) reveals polycrystalline films with columnar grains, typical for sputtered films. STEM-EDS elemental mapping indicates homogeneous incorporation of Pr in the Al$_{0.85}$Pr$_{0.15}$N film (Figure 1a), while the Al$_{0.92}$Tb$_{0.08}$N film (Figure 1b) has regions with homogeneous incorporation and regions with slight enrichment or depletion of Tb. Line profiles extracted from the EDS maps (Figure 1, bottom panels) show atomic fractions through the thickness of the films. The average surface compositions/atomic fractions of cations calculated from these line scans closely agree with those measured by EPMA (see Table S1).

The oxygen contents in the bulk of these films are approximately 5 at.%, which is typical for nitride films deposited by sputtering.[16,47,48] However, a gradual increase in oxygen towards the surface of the films suggests that a significant portion of this oxygen incorporation may have occurred post-deposition. We observe that the compositions extracted from EPMA are much closer to those extracted from the EDS line profiles near the surface than in the bulk; for example, the stoichiometry extracted from EPMA for the $Al_{0.92}Tb_{0.08}N$ film is $Al_{0.91}Tb_{0.08}N_{0.96}O_{0.17}$, while EDS yields $Al_{0.92}Tb_{0.07}N_{0.8}O_{0.15}$ near the surface. This is consistent with EPMA being more sensitive to the surface region. Table S1 presents the direct comparisons of all stoichiometries of the three films measured by EDS and EPMA. We also note that the spike in oxygen visible at the substrate interface is from the native oxide layer on Si wafers. Similar elemental mapping and line profiles for an $Al_{0.79}Tb_{0.21}N$ film are presented in Figure S3.

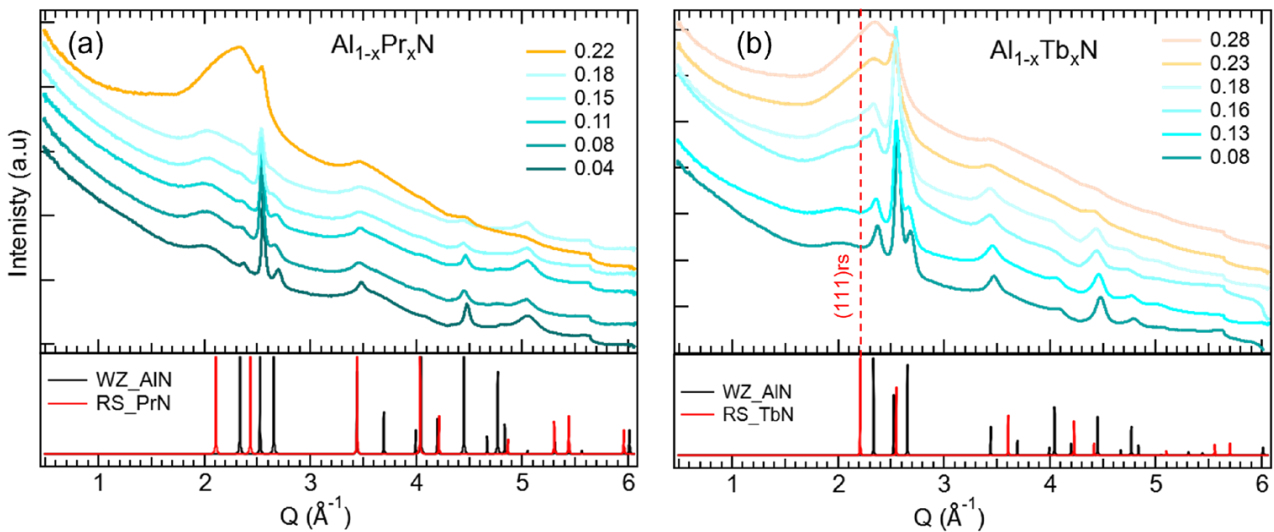

Figure 2: (a, b) Select integrated synchrotron GIWAXS data of (a) $Al_{1-x}Pr_xN$ combinatorial films with $x = 0.04$ to $0.22$, and (b) $Al_{1-x}Tb_xN$ combinatorial films with $x = 0.08$ to $0.28$. The data are shifted vertically for visual clarity.

**Crystal structure**

To understand the crystal structure as a function of RE content, we performed laboratory XRD and synchrotron GIWAXS on the $Al_{1-x}RE_xN$ ($RE$ = Pr, Tb) combinatorial films. Figure 2 displays selected integrated patterns of $Al_{1-x}Pr_xN$ and $Al_{1-x}Tb_xN$ films with a range of RE incorporation, along with the calculated patterns for the end members wz-AlN and rs-$RE$N. As seen from Figure 2a, all of the observed peaks in the $Al_{1-x}Pr_xN$ patterns up to $x \approx 0.22$ correspond to the wurtzite phase. There are no peaks that match a rocksalt phase (PrN or Al-substituted PrN). Above $x \gtrsim 0.22$, the peaks become weaker and broader, indicating significant disorder in the crystalline phase that is consistent with the presence of amorphous material (Figure S5a,b). Select data from $Al_{1-x}Tb_xN$ films (Figure 2b) show that the peaks match with reference wz-AlN peaks for approximately $x \lesssim 0.15$. However, the pattern for $x \approx 0.16$ shows an additional rs-TbN(111) peak at $Q = 2.2$ Å$^{-1}$, close to the wz-(100) peak. Additional GIWAXS patterns are shown in Figure S5e,f. Together, the data indicate that $Al_{1-x}Tb_xN$ crystallizes in a phase-pure wurtzite structure below $x \approx 0.16$ and mixed wz+rs phases for $0.16 \lesssim x \lesssim 0.28$. The onset of amorphization is observed at $x \approx 0.28$ and extends up to the highest measured Tb content of $x \approx 0.46$ (Figure S5c,d). Phase diagrams are shown in Figure 3c,d for $RE$ = Pr and Tb, respectively.

The close $d$-spacings of some of the wz-AlN and rs-$RE$N peaks and the gradual shifting of the observed peaks with $x$ complicates the phase identification with just 1D integrated patterns. Therefore, we inspect the GIWAXS 2D detector images by analyzing the tilt angles corresponding to the observed peaks to perform a detailed phase assessment. Figure 3a,b show the GIWAXS 2D detector images for select $x$-values critical for phase distinction in $Al_{1-x}RE_xN$ films. The pure wurtzite phase in $Al_{1-x}Pr_xN$ up to $x \approx 0.22$ is further corroborated by the presence of additional wurtzite peaks with their corresponding $d$-spacings and tilt angles, Figure 3a. For $Al_{1-x}Tb_xN$, the detector images in Figure 3b show that only wurtzite peaks are present for $x \approx 0.08$, and they are maintained up to $x \gtrsim 0.16$ (Figure S5d). These are the highest reported substitution levels of both $Tb^{3+}$ and $Pr^{3+}$ into wurtzite AlN. Previous reports on $Tb^{3+}$ substitution were limited to Tb contents of ~2%, while all reports on $Pr^{3+}$ involve ion implantation with concentrations similar to that of $Tb^{3+}$. In $Al_{1-x}Tb_xN$, a rs-(111) peak along with the wurtzite peaks confirms mixed wz+rs phases between $x \approx 0.16$ and $x \approx 0.28$ (Figures 1b, S5f). The amorphous nature of $Al_{1-x}Tb_xN$ films for compositions $x \gtrsim 0.28$ is confirmed by 2D detector images in Figure S5c,d.

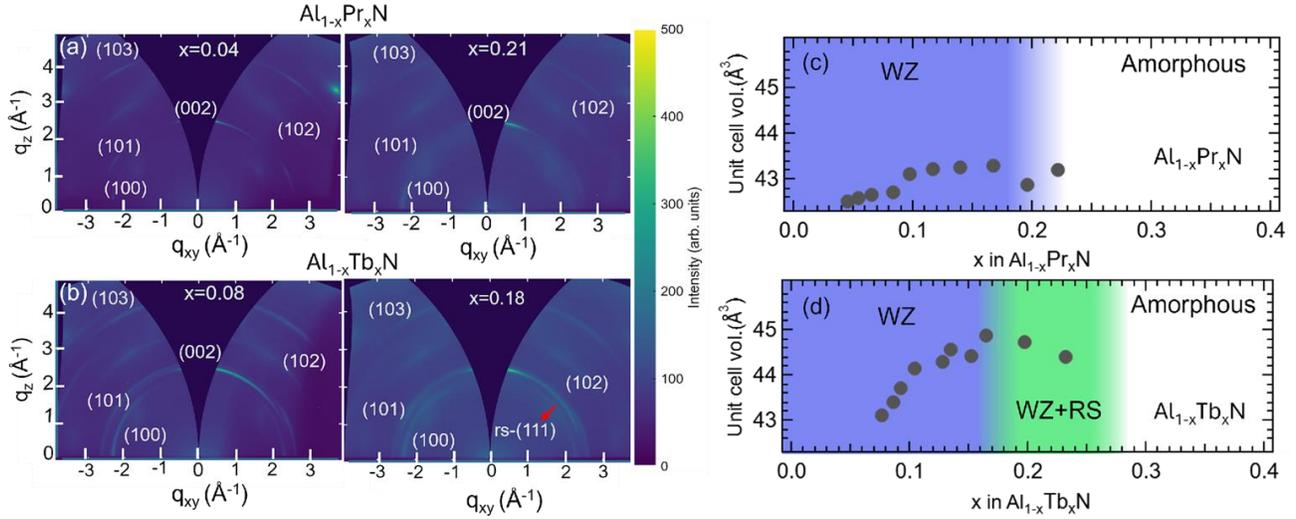

*Figure 3: (a,b) GIWAXS 2D detector images of $Al_{1-x}RE_xN$ at several x-values. (a) 2D detector images of $Al_{0.96}Pr_{0.04}N$ and $Al_{0.79}Pr_{0.21}N$ indexed to a wurtzite phase. (b) 2D detector images of $Al_{0.92}Tb_{0.08}N$ indexed to a wurtzite phase and $Al_{0.82}Tb_{0.18}N$ indexed to both wurtzite and rocksalt phases. (c,d) Evolution of unit cell volumes with substitution of (c) $Al_{1-x}Pr_xN$ and (d) $Al_{1-x}Tb_xN$ films. The phases present are marked.*

The detector images for both $RE$ = Pr and Tb films in the wurtzite phase display texturing consistent with semi-oriented films, similar to our previous work on $Al_{1-x}Gd_xN$ films.[16] Laboratory XRD data collected with a much smaller area detector reveal only the wz-(002) peak in the phase-pure wurtzite region (Figure S4a,b), which would be consistent with *c*-axis texture, as is often observed for AlN-base wurtzite films.[49,50] It is worth noting that lab XRD measurements along the surface normal may miss important crystallographic details such as off-axis diffraction and in-plane lattice parameters.

In both $Al_{1-x}RE_xN$ ($RE$ = Pr, Tb), the obvious shifting of the wurtzite peaks towards lower angles with increasing *x* (Figure 2) implies that the lattice parameters increase upon substituting RE cations onto the Al sites. To quantify this, LeBail fits were performed on the integrated GIWAXS data to extract the lattice parameters and unit cell volumes for the phases present (Figure S6a,b). The lattice parameters (Figure S7) and unit cell volume (Figure 3c,d) increase with *x* while the films maintain a phase-pure wurtzite structure. This trend is consistent with successful substitution of $RE^{3+}$ cations in wz-AlN and with prior work on $Al_{1-x}Gd_xN$ films.[16] Interestingly, we observe that the volume increase of phase-pure wurtzite $Al_{1-x}Tb_xN$ is larger than that of phase-pure wurtzite

$Al_{1-x}Pr_xN$ even though $Tb^{3+}$ is smaller than $Pr^{3+}$. The different volume increase may be caused by oxygen impurities observed by EMPA or other defects that are not the focus of this study. The trend of increasing unit cell volume in $Al_{1-x}Tb_xN$ deviates once the film changes from phase-pure wurtzite to mixed wz+rs phases, Figure 3d. In this mixed-phase regime of $Al_{1-x}Tb_xN$, the data were fitted with both wz and rs phases. The decrease in unit cell volume could imply that the $Tb^{3+}$ cations may not have substituted to Al sites once the mixed phase appears despite the wz-c and rs lattice constants increase within this range, Figure S7b. A decrease in $a$-lattice constants for the mixed phase region is observed, which dominates the unit cell volume.

The films investigated via STEM-EDS ($Al_{0.85}Pr_{0.15}N$ and $Al_{0.92}Tb_{0.08}N$) were further investigated with SAED and STEM-DF imaging, shown in Figure S8. The SAED data for $Al_{0.85}Pr_{0.15}N$ displays a diffuse ring at $d \approx 2.84$ Å, consistent with the GIWAXS data. Rings for additional wurtzite peaks are too weak to be observed, consistent with the high oxygen content of this sample. STEM-BF imaging of $Al_{0.92}Tb_{0.08}N$ reveals textured polycrystalline microstructures with columnar grains with grain sizes of approximately ~5 nm (Figure S8b). Fast Fourier transforms (FFTs) from two regions of interest are consistent with a wurtzite phase. It is evident from the STEM-BF image and FFTs patterns that some regions between the grains appear to be amorphous, and the degree of crystallinity varies throughout the film. STEM-BF imaging of $Al_{0.79}Tb_{0.21}N$ (Figure S3b) reveals a textured polycrystalline film, and several FFTs are consistent with a wurtzite-dominated phase, despite GIWAXS data showing wz+rs phases for $x \approx 0.21$ (Figure 3b,d). This may be due to microscopic segregation of wz and rs phases.

The fact that a wurtzite structure is maintained with incorporation of Pr and Tb into AlN up to approximately $x \approx 0.2$ yields similar results to our previous work on $Al_{1-x}Gd_xN$, which found Gd incorporation up to $x \approx 0.24$.[16] This is also consistent with calculations of the critical composition $x_c$ performed previously for $Al_{1-x}RE_xN$ ($RE$ = Pr, Tb), finding similar behavior to $Al_{1-x}Gd_xN$.[16] The slightly lower incorporation in the current work may be due to less optimization of growth parameters performed in the current study compared to the $Al_{1-x}Gd_xN$ work, and possibly to the less nitrogen-rich growth environment employed here. We hypothesize that the observation of a rocksalt phase in $Al_{1-x}Tb_xN$ (but not $Al_{1-x}Pr_xN$ or $Al_{1-x}Gd_xN$) might arise from its slightly lower electronegativity (and therefore less ionic bonds) compared to $Pr^{3+}$ and $Gd^{3+}$, as higher bond ionicity was shown to stabilize $Al_{1-x}Sc_xN$.[51]

**Optoelectronic properties**

To investigate the bandgap change as a function of RE composition, we performed spectroscopic ellipsometry (SE) measurements across the synthesized composition gradients of $Al_{1-x}RE_xN$ ($RE$ = Pr, Tb). Absorption coefficients ($\alpha$) extracted from the $Al_{1-x}Pr_xN$ and $Al_{1-x}Tb_xN$ films are plotted in Figure 4a,b, respectively. To avoid the errors intrinsic to Tauc analysis, we define an optical absorption cutoff as the energy when $\alpha = 5\times10^4$ cm$^{-1}$ ($E_{\alpha=5\times10^4}$),[16,52] shown as the dashed line in Figure 4a. The extracted $E_{\alpha=5\times10^4}$ values, which should follow the same trend as the bandgap, are plotted as a function of RE incorporation in Figure 4c,d, along with both direct and indirect bandgaps calculated by the GW method for wurtzite-type $Al_{1-x}RE_xN$. The calculated bandgaps fall off quickly with RE incorporation, as expected.

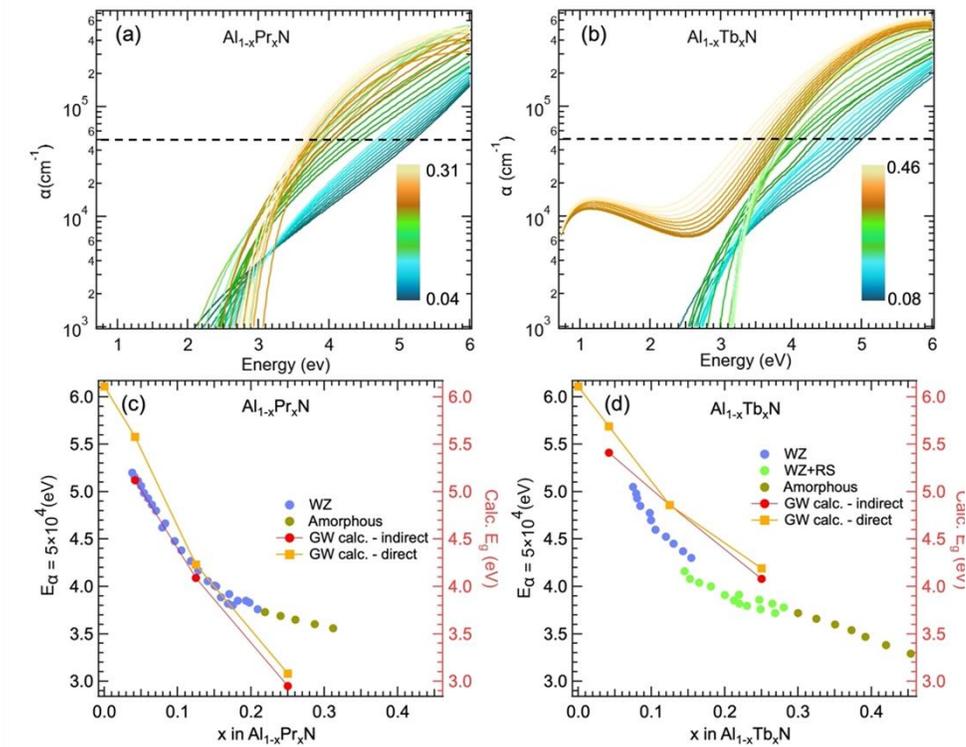

*Figure 4: (a,b) Absorption coefficients ($\alpha$) of all the compositions of combinatorial (a) $Al_{1-x}Pr_xN$ and (b) $Al_{1-x}Tb_xN$ thin films extracted from spectroscopic ellipsometry measurements. (c,d) The energy corresponding to $\alpha= 5\times10^4$ cm$^{-1}$ ($E_{\alpha=5\times10^4}$, shown by the dashed lines in a,b) and GW bandgaps as a function of x in (c) $Al_{1-x}Pr_xN$ and (d) $Al_{1-x}Tb_xN$.*

Figure 4c,d show that the experimental $E_{\alpha=5\times10^4}$ extracted for both $Al_{1-x}Pr_xN$ and $Al_{1-x}Tb_xN$ decrease with increasing RE substitution, similar to the trend measured previously for $Al_{1-x}Gd_xN$[16]. For both materials, the $E_{\alpha=5\times10^4}$ values decrease monotonically while the films maintain a pure wurtzite phase. In $Al_{1-x}Pr_xN$, the slope of the $E_{\alpha=5\times10^4}$ values changes when the film transitions from the wurtzite phase to amorphous (Figure 4c). In $Al_{1-x}Tb_xN$, the slope similarly changes between the wurtzite phase, the wz+rs region, and the amorphous region (Figure 4d). Together, the data suggest that the slope of the absorption onset, which is proportional to the bandgap, is dependent on the phases present.

Significant sub-gap absorption is visible in the $Al_{1-x}Tb_xN$ samples when $x \gtrsim 0.28$, which corresponds to the amorphous region. We hypothesize this might be due either to a mixture of higher-bandgap wz-like (i.e., tetrahedral) and lower-bandgap rs-like (i.e., octahedral) local coordination in the amorphous phase or perhaps to RE-rich regions, which would have a lower bandgap. Ellipsometry measurements in $Al_{1-x}Gd_xN$, which did not exhibit a crystalline rs phase at compositions up to $x \approx 0.24$, also did not exhibit this level of sub-gap absorption in the amorphous region, consistent with this hypothesis.

**Luminescence properties**

To study the luminescence properties of these alloys, CL measurements were performed at room temperature on our sputtered $Al_{1-x}Pr_xN$ and $Al_{1-x}Tb_xN$ films for a range of $x$ values, Figure 5. The CL spectra of $Al_{1-x}Pr_xN$ films show three main emission lines (**1**, **2**, and **3**) ranging from blue/green (~527 nm) to red (~653 nm), consistent with the previously reported luminescence spectra for $Pr^{3+}$ cations within III-nitride hosts.[20,22,53] Following previous assignments, the stronger line at 526 nm (**1**) corresponds to the intra 4f-shell $^3P_1 \rightarrow {^3H_5}$ transition, and the weaker lines (**2** and **3**) arise from the $^3P_0 \rightarrow {^3H_5}$ and $^3P_0 \rightarrow {^3F_2}$ transitions, Figure 5b. The observed emission lines are consistent with the reference spectra by ±5 nm and are tabulated in Table S2. It was previously reported that $Pr^{3+}$ cations implanted in GaN exhibited an intense red emission line at 650 nm[54,55] and that $Pr^{3+}$ in AlN in an amorphous phase yielded an intense green line at 526 nm along with reasonable intensity in the red region that increased in the crystalline phase[20,53]. These previous nitride hosts are MBE-, sputtered-, and MOCVD-deposited films with implanted $Pr^{3+}$ at low concentrations (~$10^{14}$-$10^{15}$ cm$^{-2}$); we note that structures and phase confirmations are not presented.[22,53]. It is interesting that our crystalline wurtzite films show dominant emission in the green; the red emission is significantly suppressed, in apparent contrast to earlier results.

The gradual decrease in the CL intensity with Pr incorporation, shown in the inset of Figure 5a, is attributed to concentration quenching, which is commonly observed with increasing RE concentration beyond a certain level and is independent of the phase evolution across the compositional range[56]. The significant variation in the green/red emission ratio we observe is most likely associated with a higher fraction of the Pr on the Al substitutional site, as opposed to the interstitial and defect related positions that can be associated with implanted films.

Figure 5c shows the CL spectra for $Al_{1-x}Tb_xN$ films; multiple emission lines occur within the visible region resulting from intra 4f-shell transitions of $Tb^{3+}$. The green region comprises the strongest emission line at 554 nm (**2**) and other relatively strong lines at 492 and 589 nm (**1** and **3**), while emission lines in the yellow region (**4** and others at higher wavelengths) are weaker. All observed peaks are consistent with the previously reported spectra and characteristic emissions mediated by $Tb^{3+}$ and are tabulated in Table S2, along with the transitions from which they originate.[23,24,57,58] We note that the previously reported $^5D_3$ to $^7F_J$ (J = 6, 5, 4) transitions within the blue region are not visible in our films. This is likely due to variations in cross-relaxation mechanisms that control the radiative/non-radiation ratios since the cross-relaxation process (see Fig. 5d) dominates with increasing $Tb^{3+}$ concentrations—leading to decreased emission intensities in the blue region accompanied by an increase in emission intensities in the green region[58].

The intensity of the strongest peak (**2**) shows a significant drop at higher Tb concentrations ($x \gtrsim 0.15$), which we attribute to concentration quenching (inset of Figure 5c). Interestingly, we observe reasonable intensities of the green emission lines (**1**, **2**, and **3**) in our films even at higher $Tb^{3+}$ concentrations ($x \approx 0.08$) compared to previous work, in which those lines were almost quenched at concentrations that fully suppressed the $^5D_3$ to $^7F_J$ peaks ($x \gtrsim 0.01$).[19,25,59] Our data are also in contrast to a previously reported $Al_{1-x}Tb_xN$ film with $x \lesssim 0.11$, which showed quenching of green ($^5D_4 \rightarrow ^7F_J$) emission lines above $x \approx 0.05$.[26]

In this family of materials, oxygen defects were reported to produce a broad emission peak in the UV centered around ~380 nm, arising from $(V_{Al}-O_N)^{-/2-}$ defect complexes where three oxygen impurity defects ($O_N$) create one Al vacancy ($V_{Al}$).[60,61] However, we do not observe a UV peak in either $Al_{1-x}Pr_xN$ or $Al_{1-x}Tb_xN$ films across the whole substitution range ($x$). This suggests a different form of the primary oxygen-related defect(s) in these films and may be worthy of future study. We note that the relationship between synthesis method (ion implantation, sputtering, bulk synthesis, etc.), level of crystallinity, crystalline phases present (wz, rs, etc.), $RE^{3+}$ cation position within the matrix (substitutional or interstitial), type and level of defects, and emission is not yet fully understood and is likely complex[25,62,63]. We hypothesize that stabilizing the wurtzite phase in $Al_{1-x}RE_xN$ up to higher RE composition could offer an additional knob to selectively tailor the emission lines and enhance their relative intensities, especially in co-doped films.

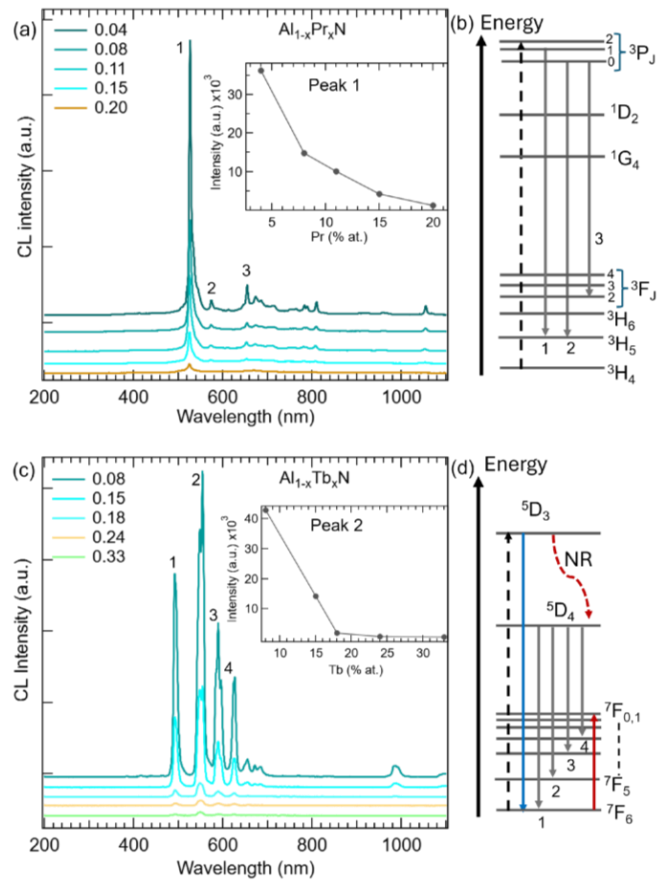

*Figure 5: Room temperature CL measurements on (a) $Al_{1-x}Pr_xN$ and (b) $Al_{1-x}Tb_xN$ films for various concentrations. Spectra are shifted vertically for clarity. Schematics of electronic transitions corresponding to the observed emission lines in (b) $Al_{1-x}Pr_xN$ and (d) $Al_{1-x}Tb_xN$ films. The cross-relaxation process is shown with red arrows in (d). Insets of panels (a,c): Intensities of the strongest CL emission line as a function of $RE^{3+}$ composition (x).*

## Conclusions

Expanding studies of metal-substituted AlN films to rare earth cation substitution offers the potential to unlock novel functionalities. Here, we have successfully synthesized a series of Al$_{1-x}$RE$_x$N (RE = Pr, Tb) thin films across a broad compositional space using rf co-sputtering. Using synchrotron GIWAXS measurements, we confirmed that Al$_{1-x}$Pr$_x$N films maintained a wurtzite structure for $x \lesssim 0.22$ and become amorphous for $x \gtrsim 0.22$. Al$_{1-x}$Tb$_x$N films were phase-pure wurtzite for $x \lesssim 0.16$; they phase segregated to mixed wurtzite and rocksalt phases for $0.16 \lesssim x \lesssim 0.28$, and they became amorphous at $x \gtrsim 0.28$. A monotonic increase in the unit cell volume with $x$ within the wurtzite phase suggests the substitution of RE$^{3+}$ onto the Al sites. This is the highest substitution of Pr$^{3+}$ and Tb$^{3+}$ into phase-pure wurtzite III-nitrides that has been reported to date, and it is comparable to the level of RE incorporation reported in our previous work on Al$_{1-x}$Gd$_x$N.

Ellipsometry measurements show a gradual decrease in the absorption onset of the films with increasing $x$, confirming theoretical predictions of the decrease in bandgap with alloying. The presence of Pr$^{3+}$ and Tb$^{3+}$ in the respective films is also confirmed by their signature emission lines from cathodoluminescence measurements. At these higher levels of incorporation for Pr, Tb and Gd in AlN, there is increased opportunity to tailor the relative intensity of luminescence across a broad range of the visible spectrum, especially with the future potential of co-doping. The demonstration of high-throughput synthesis along with compositional space-phases-properties relationship studies of this work can help establish the design rules for a broad range of AlN-based thin films.

## Author contributions

**Binod Paudel**: Investigation, Writing – original draft. **John S. Mangum**: Investigation, Writing – review & editing. **Christopher L. Rom**: Investigation, Writing – review & editing. **Kingsley Egbo**: Investigation. **Cheng-Wei Lee**: Investigation, Writing – review & editing. **Harvey Guthrey**: Investigation. **Sean Allen**: Investigation. **Nancy M. Haegel**: Funding acquisition, Project administration, Resources, Supervision, Writing – review & editing. **Keisuke Yazawa**: Supervision, Writing – review & editing. **Geoff L. Brennecka**: Funding acquisition, Project administration, Supervision, Writing – review & editing. **Rebecca W. Smaha**: Conceptualization, Supervision, Writing – review & editing.

## Conflicts of interest

There are no conflicts to declare.

## Data availability

Data that support the findings of this study are openly available in the National Renewable Energy Laboratory (NREL) high-throughput experimental materials database at https://htem.nrel.gov. Additional data that support the findings of this study are available from the corresponding author upon reasonable request.

## Acknowledgements

This work was authored by the National Renewable Energy Laboratory, operated by Alliance for Sustainable Energy, LLC, for the U.S. Department of Energy (DOE) under Contract No. DE-AC36-08GO28308. The U.S. Department of Energy Office of Science provided funding for this research collaboratively from the Office of Basic Energy Sciences, Division of Materials Science, and the Advanced Scientific Computing Research (ASCR) program. Use of the Stanford Synchrotron Radiation Lightsource, SLAC National Accelerator Laboratory, was supported by the U.S. Department of Energy, Office of Science, Office of Basic Energy Sciences under Contract No.DE-AC0276SF00515. This work also used computational resources sponsored by the Department of Energy's Office of Energy Efficiency and Renewable Energy, located at NREL. The authors thank Julie Chouinard for EPMA, Patrick Walker for FIB liftouts, EAG Eurofins for additional TEM measurements, and Nick Strange for assistance with GIWAXS measurements at SSRL. The views expressed in the article do not necessarily represent the views of the DOE or the U.S. Government.